\documentclass[sn-mathphys-num]{sn-jnl}

\usepackage{graphicx}%
\usepackage{multirow}%
\usepackage{amsmath,amssymb,amsfonts}%
\usepackage{amsthm}%
\usepackage{mathrsfs}%
\usepackage[title]{appendix}%
\usepackage{xcolor}%
\usepackage{textcomp}%
\usepackage{manyfoot}%
\usepackage{booktabs}%
\usepackage{algorithm}%
\usepackage{algorithmicx}%
\usepackage{algpseudocode}%
\usepackage{listings}%

\theoremstyle{thmstyleone}%
\theoremstyle{thmstyletwo}%

\theoremstyle{thmstylethree}%

\begin{document}

\title{Machine-Learning-Based Method for Goodness-of-Fit Test in Amplitude Analysis}


\author*[1,2]{\fnm{Guoyi} \sur{Hou}}\email{hougy@ihep.ac.cn}

\author[1,2]{\fnm{Beijiang} \sur{Liu}}\email{liubj@ihep.ac.cn}

\affil*[1]{\orgname{Institute of High Energy Physics}, \orgaddress{\city{Beijing}, \postcode{100049}, \country{People's Republic of China}}}

\affil[2]{\orgname{University of Chinese Academy of Sciences}, \orgaddress{\city{Beijing}, \postcode{100049}, \country{People's Republic of China}}}

 \abstract{\textbf{Purpose:} Amplitude analysis is a pivotal tool in hadron spectroscopy, fundamentally involving a series of likelihood fits to multi-dimensional experimental distributions. While robust goodness-of-fit tests exist for low-dimensional scenarios, evaluating goodness-of-fit in amplitude analysis remains challenging.
 
 \textbf{Methods:} We propose a machine-learning approach using anomaly detection for goodness-of-fit assessment in amplitude analysis. Our method employs a classifier to identify discrepancies between data and fit results in multi-dimensional phase space.
 
 \textbf{Results and Conclusion:} Using Monte Carlo simulations of $J/\psi\to\gamma \pi^+\pi^-\pi^0\pi^0$ decays, we demonstrate that this method detects contributions from an additional resonance with a signal strength of 1\%. The detection power is sufficient for practical amplitude analyses, where contributions with fit fractions larger than 1\% are typically included in the nominal fit. This approach shows promise for amplitude analyses of multi-body processes.
}




\maketitle

\section{Introduction}\label{sec1}

The strong force, governed by Quantum Chromodynamics (QCD), regulates the interactions between quarks and gluons, the fundamental constituents responsible for most visible mass in the universe. A central challenge in QCD is elucidating the complex hadron spectrum in terms of quark and gluon degrees of freedom. Experiments such as BESIII~\cite{BES3}, LHCb~\cite{LHCb}, and Belle II~\cite{BelleII} have extensively explored both ordinary and exotic hadrons. Due to the short-lived nature of these states, resonance properties must be inferred from the kinematic distributions of their decay products.

The primary task of experimental hadron spectroscopy is to systematically map out resonances and determine their properties---such as mass, width, spin-parity, and partial decay widths---with high sensitivity and accuracy. However, extracting resonance properties from data is non-trivial: resonances are often broad, overlapping, or obscured by background. High statistical precision and sophisticated analysis methods, particularly amplitude analysis techniques, are essential for disentangling contributions from individual resonances and determining their quantum numbers~\cite{PWA}.

Amplitude analysis involves unbinned maximum likelihood fits to high-dimensional distributions of decay products, accounting for full kinematic correlations. Goodness-of-fit tests are critical for assessing the agreement between the fit model and data. While robust methods like $\chi^2$ or Kolmogorov-Smirnov tests exist for low dimensions, they lose power in high-dimensional spaces due to the "curse of dimensionality", as finite samples become sparse in high-dimensional spaces~\cite{Bellman}.
For an n-body final state, $3n-4$ kinematic variables are required to fully define the decay kinematics. Therefore, goodness-of-fit testing in amplitude analysis is a challenging issue. 
 
A goodness-of-fit test can be effectively interpreted as a problem of determining whether two data samples are drawn from the same (unknown) statistical distribution.
Machine learning (ML) classification was first proposed for multivariate goodness-of-fit and two-sample testing~\cite{Friedman}. 
Various ML-based anomaly detection methods have been explored in high-energy physics to search for new physics beyond the Standard Model~\cite{AD1,AD2,AD3,AD4,AD5}. 
The goal is to identify discrepancies between experimental data, which may contain a mixture of background and potential signal events, and a reference background dataset.
These binary classification procedures can be adapted for high-dimensional two-sample testing, where classifier outputs are used to perform univariate tests.

In this work, we employ ML-based anomaly detection to evaluate goodness-of-fit in amplitude analyses.
Anomalous data behavior is identified by comparing high-dimensional distributions between experimental data and Monte Carlo (MC) samples generated according to the amplitude analysis results.
We construct a classifier using XGBoost~\cite{XGBoost} to detect localized anomalies in high-dimensional data. Statistical tests are developed using the probabilistic classifier's output. The objective is to determine whether there is any statistically significant difference between the distribution of data and the amplitude analysis results. We demonstrate the performance of this approach using the amplitude analysis of the multi-body decay process $J/\psi\to\gamma 4\pi$ as an example. 
Additionally, goodness-of-fit is evaluated for various signal strengths and types of intermediate resonances.
The ML-based anomaly detection approach demonstrates promising performance in testing goodness-of-fit for amplitude analyses.


\section{Method}\label{sec2}

In amplitude analysis, an unbinned maximum likelihood fit of a probability density function (PDF) to an experimental dataset is performed. A common approach involves generating one-dimensional histograms, such as invariant mass and angular distributions, for the data and the projections of the fit model. These projections are obtained using an MC sample generated according to the amplitude analysis results. The $\chi^2/N_{bin}$ for each one-dimensional distribution is used to assess goodness-of-fit, where $N_{bin}$ is the number of bins in the histogram. The $\chi^{2}$ is defined as
$
\chi^{2}=\sum_{i=1}^{N_{{\rm bin}}}\frac{(n_{i}-\nu_{i})^{2}}{\nu_{i}},
$
where $N_{bin}$ is the number of bins in the histogram, $n_{i}$ and $\nu_{i}$ represents the number of events in the data and fit projections, respectively, for the $i$-th bin of the distribution. However, this approach has limitations. First, it cannot account for correlations between distributions. Second, data become sparse as the dimensionality of the histograms increases.

Goodness-of-fit tests evaluate whether observed data $X={\{x_i\}}^{N_x}_{i=1}$ matches a probability density distribution $p(x)$. The goal of a two-sample test is to compare datasets  $X$(drawn from $p(x)$) and $Y={\{y_i\}}^{N_y}_{i=1}$(drawn from $q(y)$) to test whether $p$ equals to $q$. Following the method proposed in Ref.~\cite{Friedman}, binary multivariate classification can be used in two-sample tests. $X$ and $Y$ are combined into a labeled dataset $Z={\{z_i,s_i\}}$, where the label $s_i=1$ if $z_i \in Y$ and $s_i=0$ if $z_i \in X$. An ML probabilistic classifier (e.g., boosted decision tree, neural network) is trained on $Z$ to  distinguish $X$ versus $Y$.  The trained classifier, $g(z) = P(Y|Z=z)$, is used to estimate the probability for $z_i \in X$. In analogy to a high-dimensional signal-background classification, an ML classifier can be used to discriminate between the MC sample generated from the amplitude analysis fit results and the experimental data. The classifier is trained to classify which sample an event belongs to based on the measured values of its features. The objective is to establish a hypothesis test of $H_0$(the null hypothesis that the data sample and the MC sample share the same distribution) versus $H_1$ (the alternative hypothesis that there is a deviation between the data and the MC sample). Using the output of the classifier, test statistics can be constructed with the approach in Ref.~\cite{AD4} to assess the presence of a deviation between data and the amplitude analysis fit results. 
In this work, we adopt an estimated Likelihood Ratio Test (LRT) statistic and a test based on the Area Under the Receiver Operating Characteristic (ROC) Curve (AUC). 
As defined  in Ref.~\cite{AD4}, the LRT statistic is based on the likelihood ratio of the experimental data under the null hypothesis versus the alternative hypothesis, measuring the total log-likelihood ratio for the data events. Large LRT values indicate evidence against the null hypothesis. Alternatively, the AUC statistic measures the classifier's ability to distinguish between the data and MC samples. The ROC curve plots the true positive rate against the false positive rate at various classification thresholds, visualizing the classifier's discrimination performance. Under the null hypothesis, the AUC should be close to 0.5. A significantly higher AUC indicates the presence of an anomaly (a deviation between the data and MC). 

The null distribution is the distribution of the test statistic under the null hypothesis, which states that there is no difference between the data and MC. Estimating this distribution is essential for determining the significance of the observed test statistic and deciding whether to reject the null hypothesis. Since asymptotic distributions are unreliable for the test statistics of LRT or AUC when using estimated classifiers, we use the bootstrap method proposed in Ref.~\cite{AD4} to approximate the null distribution. We repeatedly resample the data and MC samples with replacement, recompute the test statistic for each resampled dataset, and use the distribution of these statistics to approximate the null distribution. The bootstrap method is flexible and does not rely on asymptotic approximations, making it suitable for high-dimensional data. To detect localized anomalies, p-values are calculated using the test statistics and their estimated null distributions. The p-value is a measure of the evidence against the null hypothesis, used to determine whether the observed data significantly deviates from the amplitude analysis fit results. It is the probability of observing a test statistic as extreme as, or more extreme than, the one observed, assuming the null hypothesis is true. A smaller p-value indicates stronger evidence against the null hypothesis. Typically, a significance level (e.g., $\alpha=0.05$) is chosen, and if the p-value is less than this level, the null hypothesis is rejected. 

\begin{figure}[!htbp]
\centering
\includegraphics[width=0.8\textwidth]{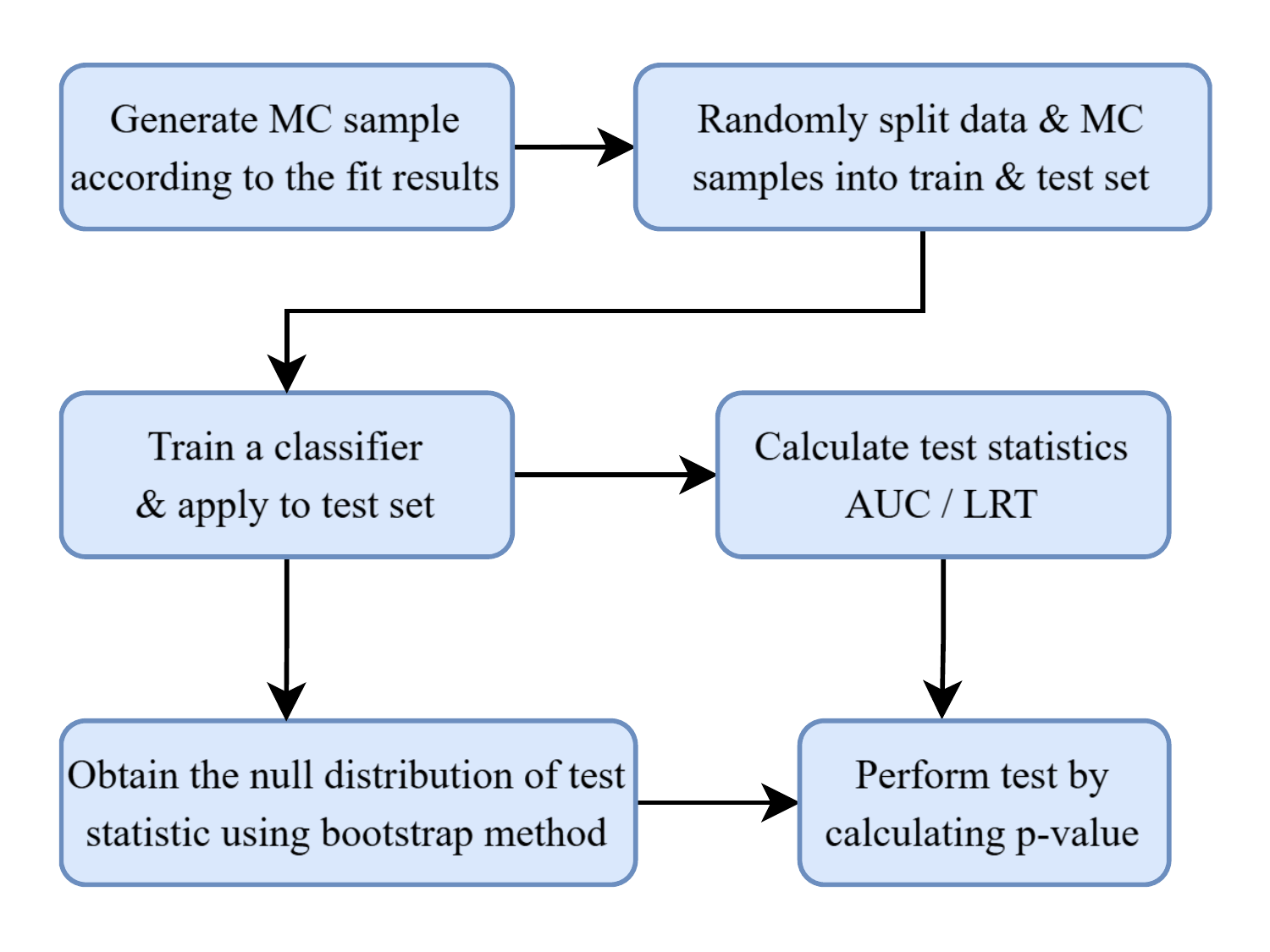}
\caption{Workflow for the goodness-of-fit test approach using anomaly detection.}\label{fig:workflow}
\end{figure}

The workflow for this goodness-of-fit test approach is illustrated in Figure~\ref{fig:workflow}. The procedures for the generation of MC sample and  the caluculation of p-value are described in Section~\ref{sec4}.

\section{Anomaly Detection Using XGBoost}\label{sec3}

Given the dimensionality and size of the datasets for amplitude analysis, XGBoost (Extreme Gradient Boosting)~\cite{XGBoost} is a suitable algorithm for constructing the probabilistic classifier in this work. XGBoost is a powerful machine learning algorithm that enhances the traditional gradient boosting framework. It combines multiple weak models, usually decision trees, into a strong predictive model by iteratively adding new trees to correct the errors of the previous ones, guided by gradient descent to minimize the loss function. XGBoost stands out for its efficiency, scalability, and robustness, incorporating features like regularization to prevent overfitting, handling of missing values, and parallelized computations for faster processing. These capabilities make it highly effective for analyzing complex datasets, including those in particle physics, where high precision and accuracy are essential.

Table~\ref{tab:hyperparam} presents the configuration of XGBoost hyperparameters. The selection of these parameters is important for the model's performance and is highly task-dependent. To ensure a robust and data-driven configuration, we did not select them manually. Instead, the optimal values were determined through an automated Bayesian Optimization process~\cite{snoek2012practical}. This process systematically searched the hyperparameter ranges, as specified in Table~\ref{tab:hyperparam}, with the objective of minimizing the `logloss` metric. 
Key hyperparameters include $\textit{eta}$ (learning rate) which controls the step size at each boosting iteration. Lower values require more trees but improve generalization. In this work, $\textit{eta}$ is optimized to 0.1. $\textit{max\_depth}$, which controls model complexity by limiting the depth of individual decision trees, is optimized to 6. $\textit{min\_child\_weight}$ determines the minimum sum of instance weights required in a child node and is optimized to 5.0, further controlling tree complexity. $\textit{gamma}$ defines the minimum loss reduction required to make a further partition on a leaf node of the tree, controlling tree pruning, and is optimized to 0.1. $\textit{subsample}$ means the subsample ratio of training instances, is optimized to 0.8. Similarly, $\textit{colsample\_bytree}$, the subsample ratio of columns when constructing each tree, is optimized to 1.0. $\textit{lambda}$, the L2 regularization term, is optimized to 4.5, while $\textit{alpha}$, the L1 regularization term, is set to 0.1.  A 10-fold cross-validation is employed to improve generalization performance and reduce overfitting. The model training of the classifier is described in the next section.

\begin{table}[!htbp]
\centering
\caption{Configuration of XGBoost hyperparameters}\label{tab:hyperparam}%
\begin{tabular}{@{}llll@{}}
\toprule
Option & Configuration & Range & Description\\
\midrule
objective & binary:logistic & - & Classification objective function \\
eval\_metric & logloss & - & Evaluation metric \\
eta & 0.1 & 0.01--0.3 & Learning rate \\
max\_depth & 6 & 3--10 & Maximum depth of the decision tree \\
min\_child\_weight & 5.0 & 1.0--10.0 & Minimum sum of instance weight \\ & & & needed in a child \\
subsample & 0.8 & 0.5--1.0 & Subsample ratio of the training instances \\
colsample\_bytree & 1.0 & 0.5--1.0 & Subsample ratio of columns when \\ & & & constructing each tree \\
lambda & 4.5 & 0.1--10.0 & L2 regularization term \\
alpha & 0.1 & 0.0--1.0 & L1 regularization term \\
gamma & 0.1 & 0.0--1.0 & Minimum loss reduction required for a split \\
\botrule
\end{tabular}
\end{table}


\section{Experiments}\label{sec4}

\subsection{Data Description}

We use the amplitude analysis of the $J/\psi\to\gamma \pi^+\pi^-\pi^0\pi^0$ process to demonstrate the performance of the goodness-of-fit test. The studies of glueballs and hybrids are crucial for gaining insights into the gluon degree of freedom of the strong interaction at low energy scales, which is essential for understanding the non-perturbative QCD. Hybrids and glueballs are expected to be abundantly produced in gluon-rich processes, such as $J/\psi$ decays. 
The BESIII experiment, with its collection of 10 billion $J/\psi$ events, provides an ideal laboratory for exploring glueballs and hybrids.  
The $J/\psi\to\gamma 4\pi$ process is of importance for the study of glueballs and hybrids. 

For a typical three-body decay, $J/\psi\to ABC$, the phase space can be completely described with three Euler angles that specify the orientation of the final system relative to the initial particle and a Dalitz plot that specifies the decay process~\cite{ParticleDataGroup:2024cfk}. 
The standard form for the Dalitz plot is defined with $M^2(AB)$ and $M^2(AC)$, where $M(AB(C))$ is the invariant mass of final state particles A and B(C).
For $J/\psi$ produced from $e^+e^-$ annihilation, the final-state particles of the $J/\psi\to\gamma 4\pi$ process can be represented in a 10-dimensional phase space, given the trivial integral over the rotation around the beam axis. An event of $J/\psi\to\gamma 4\pi$ is characterized by the measured four-momenta of the final particles, which is represented by 10 independent observables. In this work, we select invariant masses of eight pairs of final-state particles and two of the three Euler angles specifying the orientation of the final system relative to the $e^+e^-$ as the coordinates of the 10-dimensional phase space for the $J/\psi\to\gamma 4\pi$ process. 
Even though different feature designs for the classifier can be sensitive to particular anomalies, we use a general description of the multi-body phase space as input variables. This is because the specific deviations of the fit results from the data are unknown in a practical analysis. 

The kinematic observables used to represent the phase space are as follows:

\begin{itemize}
    \item $cos\theta(X)$: Cosine of polar angle of X ($4\pi$) in $J/\psi$ rest frame
    \item $\phi_X(\pi^+)$: Azimuthal angle of $\pi^+$ in X helicity frame
    \item $M(\pi^+\pi^-)$: Invariant mass of $\pi^+\pi^-$
    \item $M(\pi^+\pi^0_1)$: Invariant mass of $\pi^+$ and $\pi^0_1$, where $\pi^0_1$ denotes the $\pi^0$ with higher energy
    \item $M(\pi^+\pi^0_2)$: Invariant mass of $\pi^+$ and $\pi^0_2$, where $\pi^0_2$ denotes the $\pi^0$ with lower energy
    \item $M(\pi^-\pi^0_1)$: Invariant mass of $\pi^-$ and $\pi^0_1$
    \item $M(\pi^-\pi^0_2)$: Invariant mass of $\pi^-$ and $\pi^0_2$
    \item $M(\pi^0_1\pi^0_2)$: Invariant mass of $\pi^0_1\pi^0_2$
    \item $M(\gamma\pi^0_1)$: Invariant mass of $\gamma$ and $\pi^0_1$
    \item $M(\gamma\pi^0_2)$: Invariant mass of $\gamma$ and $\pi^0_2$
\end{itemize}

\subsection{Experiment setup}

For demonstration purposes, we consider an amplitude analysis of experimental data searching for new resonances. The underlying physics of the experimental data is described by a specific set of amplitudes with intermediate resonances. A pseudo-data sample is generated with a set of amplitudes with intermediate resonances as the ground truth. Quasi-two-body amplitudes in the sequential radiative decay processes $J/\psi\to\gamma X, X\to Y \pi, Y\to \pi Z, Z\to \pi\pi$ are constructed using the covariant tensor amplitudes described in Ref.~\cite{Zou:2002ar}. For the pseudo-data sample, the full set of amplitudes includes four subprocesses as listed in Table~\ref{tab:resonances}: $J/\psi\rightarrow\gamma \eta(1760), \eta(1760)\rightarrow\rho\rho, \rho\rightarrow\pi\pi$; $J/\psi\rightarrow\gamma f_2(2340), f_2(2340)\rightarrow\rho\rho, \rho\rightarrow\pi\pi$; $J/\psi\rightarrow\gamma f_0(2100), f_0(2100)\rightarrow\sigma\sigma, \sigma\rightarrow\pi\pi$; and $J/\psi\rightarrow\gamma f_0(2100), f_0(2100)\rightarrow\pi \mathbf{a_1(1260)}, a_1(1260)\rightarrow\rho\pi, \rho\rightarrow\pi\pi$. 

In cases where the fit model differs from the ground truth, for instance, if a subprocess is missing in the fit model, discrepancies between the data and fit results will arise. Here, we consider two distinct scenarios: Case 1, where a subprocess $J/\psi\rightarrow\gamma f_0(2100), f_0(2100)\rightarrow\pi \mathbf{a_1(1260)}, a_1(1260)\rightarrow\rho\pi, \rho\rightarrow\pi\pi$ is missing in the fit model, corresponding to a resonance in the $3\pi$ spectrum($a_1(1260)$); and Case 2, where a subprocess $J/\psi\rightarrow\gamma \mathbf{f_0(2100)}, f_0(2100)\rightarrow\pi a_1(1260), a_1(1260)\rightarrow\rho\pi, \rho\rightarrow\pi\pi$ is missing in the fit model, corresponding to a resonance in the $4\pi$ system($f_0(2100)$). It should be noted that the mass of $3\pi$ or $4\pi$ is not used as an input variable for the classifier. MC samples are generated according to the fit models. The MC sample is compared to the pseudo-data sample with the classifier to conduct the goodness-of-fit test. In all experiments, the pseudo-data set contains $5 \times 10^5$ events, each generated with a full set of four amplitudes listed in Table~\ref{tab:resonances}. 
The MC dataset also contains $5 \times 10^5$ events, each generated with a set of three amplitudes from Table~\ref{tab:resonances}. Event information at the generator level is used to produce the samples.

\begin{table}[!htbp]
    \centering
    \caption{The subprocesses with intermediate resonances included in the MC or pseudo-data set. The additional resonances are marked in bold. Case 1: A subprocess $J/\psi\rightarrow\gamma f_0(2100), f_0(2100)\rightarrow\pi \mathbf{a_1(1260)}, a_1(1260)\rightarrow\rho\pi, \rho\rightarrow\pi\pi$ is missing in the fit model, corresponding to a resonance in the $3\pi$ spectrum($a_1(1260)$). Case 2: A subprocess  $J/\psi\rightarrow\gamma \mathbf{f_0(2100)}, f_0(2100)\rightarrow\pi a_1(1260), a_1(1260)\rightarrow\rho\pi, \rho\rightarrow\pi\pi$ is missing in the fit model, corresponding to a resonance in the $4\pi$ system($f_0(2100)$). }\label{tab:resonances}%
    \begin{tabular}{@{}llll@{}}
        \toprule
        Subprocess & Pseudo-data set & MC set \\
        \midrule
        \multicolumn{3}{l}{\textbf{Case 1}} \\
        \midrule
        $J/\psi\rightarrow\gamma \eta(1760), \eta(1760)\rightarrow\rho\rho, \rho\rightarrow\pi\pi$ & $\checkmark$ & $\checkmark$ \\
        $J/\psi\rightarrow\gamma f_2(2340), f_2(2340)\rightarrow\rho\rho, \rho\rightarrow\pi\pi$ & $\checkmark$ & $\checkmark$ \\
        $J/\psi\rightarrow\gamma f_0(2100), f_0(2100)\rightarrow\sigma\sigma, \sigma\rightarrow\pi\pi$ & $\checkmark$ & $\checkmark$ \\
        $J/\psi\rightarrow\gamma f_0(2100), f_0(2100)\rightarrow\pi \mathbf{a_1(1260)}, a_1(1260)\rightarrow\rho\pi, \rho\rightarrow\pi\pi$ & $\checkmark$ & $\times$ \\
        \midrule
        \multicolumn{3}{l}{\textbf{Case 2}} \\
        \midrule
        $J/\psi\rightarrow\gamma \eta(1760), \eta(1760)\rightarrow\rho\rho, \rho\rightarrow\pi\pi$ & $\checkmark$ & $\checkmark$ \\
        $J/\psi\rightarrow\gamma f_0(1500), f_0(1500)\rightarrow\rho\rho, \sigma\rightarrow\pi\pi$ & $\checkmark$ & $\checkmark$ \\
        $J/\psi\rightarrow\gamma \eta_1(1855), \eta_1(1855)\rightarrow\pi a_1(1260), a_1(1260)\rightarrow\rho\pi, \rho\rightarrow\pi\pi$ & $\checkmark$ & $\checkmark$ \\
        $J/\psi\rightarrow\gamma \mathbf{f_0(2100)}, f_0(2100)\rightarrow\pi a_1(1260), a_1(1260)\rightarrow\rho\pi, \rho\rightarrow\pi\pi$ & $\checkmark$ & $\times$ \\
        \botrule
    \end{tabular}
\end{table}

The signal strength of the anomaly $\lambda$ is represented by the contribution of a missing subprocess in the differential cross-section using a standard definition of the fit fraction, $\frac{\sigma_X}{\sigma'}$, where $\sigma^\prime\equiv \int{\left|M(\xi) \right|^2\Phi(\xi)d\xi}$ and  $\sigma_X\equiv \int{\left|A_X(\xi_i) \right|^2\Phi(\xi)d\xi}$. 
Detection efficiency is assumed to be 100\%, as the data samples are produced at the generator level. An event is characterized by the measured four-momenta of the final particles, $\xi$. 
$\Phi(\xi)$ is the standard element of phase space, and $M(\xi) = \sum_{X} A_X(\xi)$ is the matrix element describing the decay processes from the $J/\psi$ to the final state $\gamma 4\pi$. 
$A_X(\xi)$ is the amplitude of the missing subprocess in the fit model corresponding to the intermediate resonance X. 
We generate pseudo-data sets with various signal strengths $\lambda=0.001$ , $\lambda=0.003$, $\lambda=0.005$, and up to $\lambda=0.1$(the contribution of the missing subprocess in the differential cross-section is 0.1\%, 0.5\%, and up to 10\%).
A pseudo-data set with $\lambda=0$ is also generated to verify the null case, where the pseudo-data and MC share the same model.

We perform anomaly detection using the classifier described in Sect.~\ref{sec3}. 
The pseudo-data and MC samples are randomly split into $2.5 \times 10^5$ events for training and $2.5 \times 10^5$ events for testing. 
For each signal strength case, 50 random samplings of the data are performed. For each bootstrap method, 1000 bootstrap cycles are conducted.

The p-value is calculated based on the observed test statistic and its distribution under the null hypothesis. First, the test statistic is calculated from the dataset. The null distribution that the test statistic follows under the null hypothesis is approximated using the bootstrap method. Then, the p-value can be obtained from a one-tailed hypothesis test, in which the p-value is the probability of observing a test statistic equal to or larger than the observed test statistic.

\begin{figure}[!htbp]
\centering
\includegraphics[width=1.0\textwidth]{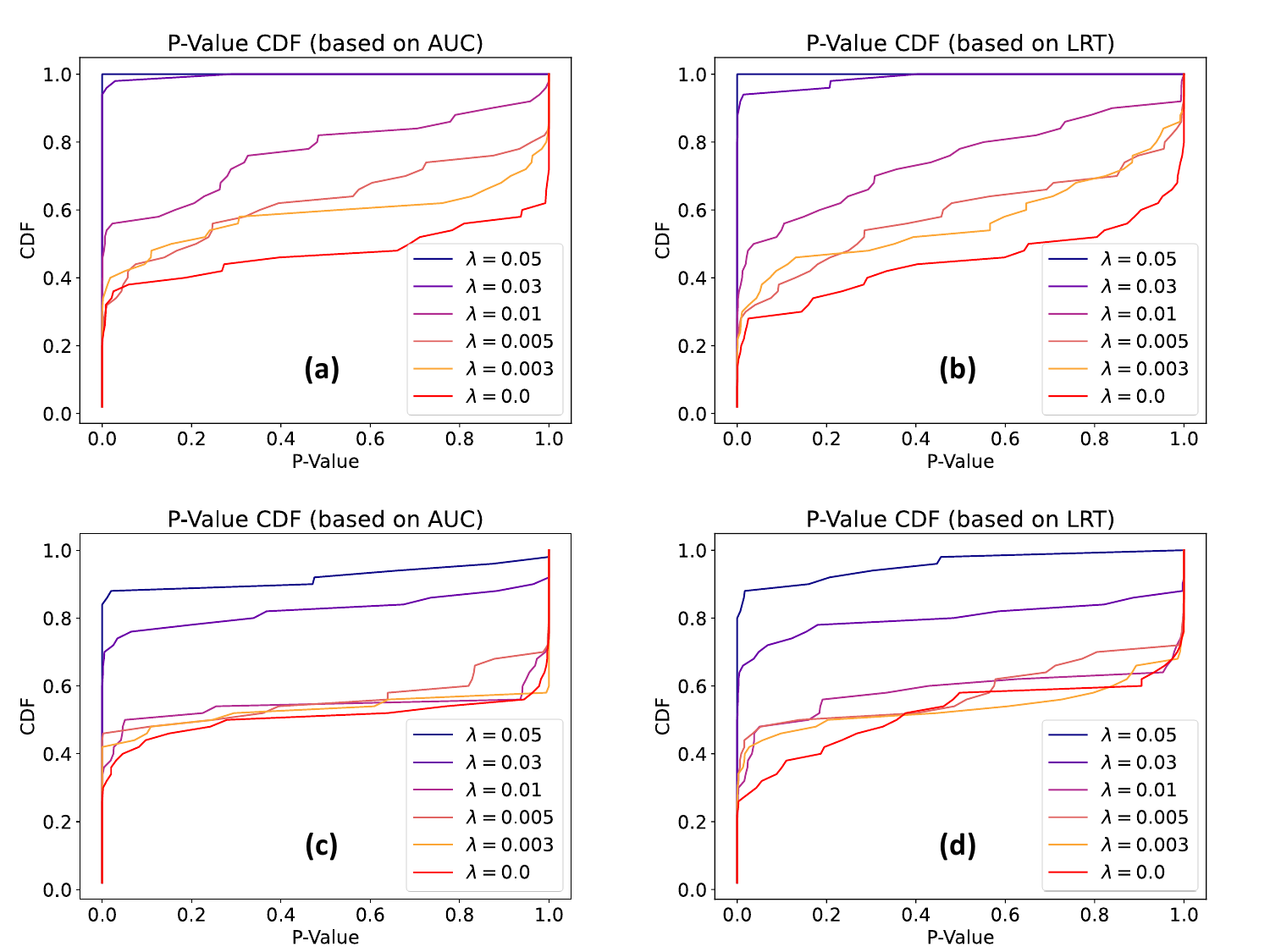}
\caption{Empirical p-value distributions under varying signal strengths for additional resonances in (a)(b) Case 1 and (c)(d) Case 2. (a)(c) Using AUC statistic; (b)(d) Using LRT statistic. The red line ($\lambda=0$) represents the null case. }\label{fig:cdf_pvalues}
\end{figure}

Figure~\ref{fig:cdf_pvalues}  shows the empirical distributions of the p-values produced by the classifier tests for different signal strengths. The cumulative distribution functions (CDFs) in the null case are approximately uniformly distributed, indicating good control of false-positive errors. The empirical CDF of p-values is used to visualize the distribution of p-values obtained from a set of statistical tests. It helps in assessing whether the p-values are consistent with the null hypothesis (i.e., the two samples share the same distribution). The X-axis is p-value and the Y-axis is Cumulative Proportion, which is the fraction of p-values that are less than or equal to p. In the context of this demonstration with the amplitude analysis of $J/\psi\to\gamma 4\pi$, the power of the test is the probability that the test correctly identifies the presence of an anomaly (a deviation between the data and MC). At a confidence level of 95\%($\alpha$=0.05) for all tests, Table~\ref{tab:detection_rate} shows the rate at which each test rejects the null hypothesis that no signal is present. It is shown that the classifier is able to detect the contribution from an additional resonance with a signal strength of 0.01 in the tests of Case 1 and Case 2. The power of the test is sufficient for practical amplitude analyses, where contributions with fit fractions larger than 1\% are typically included in the nominal fit.

\begin{figure}[!htbp]
\centering
\includegraphics[width=0.9\textwidth]{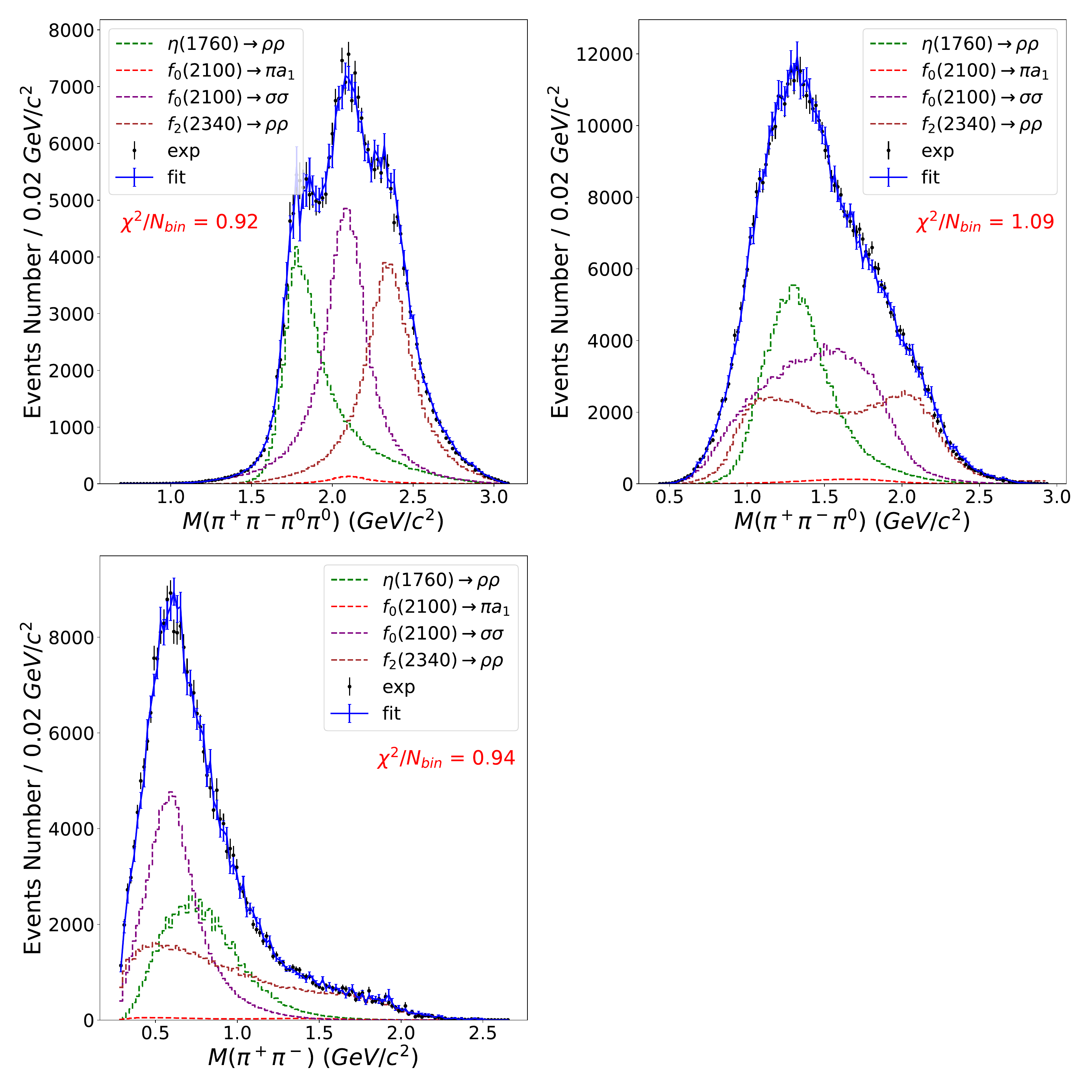}
\caption{Invariant mass distributions of $4\pi$, $3\pi$ and $2\pi$ for the data and the projects of amplitude analysis with an additional resonance in $3\pi$ ($\lambda = 0.01$). Black dots with error bars represent "exp" (experiment data) and the blue lines represent "fit" (fit results). Those dashed lines represent intensity of each component in the PWA model of "exp". The green lines represent $J/\psi\rightarrow\gamma \eta(1760), \eta(1760)\rightarrow\rho\rho, \rho\rightarrow\pi\pi$. The brown lines represent $J/\psi\rightarrow\gamma f_2(2340), f_2(2340)\rightarrow\rho\rho, \rho\rightarrow\pi\pi$. The purple lines represent $J/\psi\rightarrow\gamma f_0(2100), f_0(2100)\rightarrow\sigma\sigma, \sigma\rightarrow\pi\pi$. The red lines represent the anomaly signal $J/\psi\rightarrow\gamma f_0(2100), f_0(2100)\rightarrow\pi a_1(1260), a_1(1260)\rightarrow\rho\pi, \rho\rightarrow\pi\pi$. The mass distribution of $3\pi$ contains two $\pi^0$ combinations, meaning that each event contains 2 entries in the histogram. }\label{fig:mass_distribution}
\end{figure}

Figure~\ref{fig:mass_distribution} shows the invariant mass distributions of $4\pi$, $3\pi$ and $2\pi$ for the data and the projects of amplitude analysis for the fit model in Case 1 with the signal strength of anomaly $\lambda = 0.01$. 
Black dots with error bars represent "exp" (experiment data) and the blue lines represent "fit" (fit results). Those dashed lines represent intensity of each component in the PWA model of "exp". The green lines represent $J/\psi\rightarrow\gamma \eta(1760), \eta(1760)\rightarrow\rho\rho, \rho\rightarrow\pi\pi$. The brown lines represent $J/\psi\rightarrow\gamma f_2(2340), f_2(2340)\rightarrow\rho\rho, \rho\rightarrow\pi\pi$. The purple lines represent $J/\psi\rightarrow\gamma f_0(2100), f_0(2100)\rightarrow\sigma\sigma, \sigma\rightarrow\pi\pi$. The red lines represent the anomaly signal $J/\psi\rightarrow\gamma f_0(2100), f_0(2100)\rightarrow\pi a_1(1260), a_1(1260)\rightarrow\rho\pi, \rho\rightarrow\pi\pi$. The mass distribution of $3\pi$ contains two $\pi^0$ combinations, meaning that each event contains 2 entries in the histogram. 
The $\chi^2/N_{bin}$ value for the one-dimensional histogram is displayed on each figure. The classifier-based goodness-of-test is able to detect the anomaly, in the case that a deviation between the data and MC is difficult to be detected with the traditional $\chi^2$ test.  

\begin{table}[!htbp]
\centering
\caption{Rejection rate of the null hypothesis (no signal) for different signal strengths.}\label{tab:detection_rate}
\begin{tabular}{llcccccc}
\toprule%
& & \multicolumn{6}{c}{Signal Strength ($\lambda$)} \\
\cmidrule{3-8}
 & Test Statistic & 0.1 & 0.05 & 0.03 & 0.01 & 0.005 & 0 \\
\hline
Case 1 & AUC & 100 & 100 & 98 & 56 & 38 & 36 \\
 & LRT & 100 & 100 & 94 & 50 & 32 & 28 \\
\hline
Case 2 & AUC & 100 & 88 & 74 & 48 & 46 & 40 \\
 & LRT & 100 & 88 & 70 & 48 & 46 & 30 \\
\botrule
\end{tabular}
\end{table}

\section{Conclusion and Outlook}\label{sec5}

This study adopts a machine-learning-based approach for goodness-of-fit testing in high-dimensional amplitude analyses, addressing limitations of traditional methods like $\chi^2$ tests, which often fail in complex, multi-dimensional settings. By employing a probabilistic classifier (e.g., XGBoost), we have demonstrated a robust method to identify deviations between experimental data and amplitude analysis fit results. Using the $J/\psi\to\gamma 4\pi$ decay as a test case, our method successfully detected contributions from additional resonances with signal strengths as low as 1\%. 
As an alternative to the traditional $\chi^2$ methods commonly used in BES and LHC experiments, our approach employs a classifier to distinguish between data and Monte Carlo samples generated from the fit model. The classifier's output is used to construct test statistics, and the null distribution is estimated using a bootstrap method, allowing us to calculate p-values and quantitatively assess the evidence against the null hypothesis. This method provides a powerful tool for identifying deviations in high-dimensional phase spaces, enhancing the reliability of amplitude analyses.
Future work will focus on several key areas. First, incorporating systematic uncertainties, such as detection efficiency and background modeling, is essential for robust goodness-of-fit assessments. Second, applying our method to other multi-body decay processes and datasets will further validate its effectiveness and general applicability. Finally, exploring advanced machine learning techniques, like deep learning models, could enhance sensitivity and accuracy in detecting subtle deviations.
In summary, our machine-learning-based anomaly detection method represents a significant advancement in goodness-of-fit testing for high-dimensional amplitude analyses. By addressing key limitations of traditional methods, this approach offers a powerful tool for identifying deviations and improving the accuracy of amplitude analyses in hadron spectroscopy. Future work will refine this method and explore its broader applications in particle physics.

~\\
\noindent\textbf{Acknowledgements} This work is supported in part by National Natural Science Foundation of China (NSFC) under Contracts Nos.12235017, 12361141819; the Chinese Academy of Sciences Youth Team Program under Contract No. YSBR-101; the Strategic Priority Research Program of Chinese Academy of Sciences under Grant XDA0480600.

\section*{Declaration}
\textbf{Conflict of interest} On behalf of all authors, the corresponding author states that there is no conflict of interest.

%
%


\end{document}